\documentclass{PoS}

\usepackage{amsmath}
\usepackage{slashed}

\def\LL{\left\langle}   
\def\RR{\right\rangle}  
\def\PAR#1#2{ {{\partial #1}\over{\partial #2}} }
\def\PARTWO#1#2{ {{\partial^2 #1}\over{\partial #2}^2} }

\newcommand{\BE}{\begin{displaymath}}
\newcommand{\EE}{\end{displaymath}}
\newcommand{\BNE}{\begin{equation}}
\newcommand{\ENE}{\end{equation}}
\newcommand{\BEA}{\begin{eqnarray}}
\newcommand{\EEA}{\nonumber\end{eqnarray}}

\newcommand{\Tr}{{\rm Tr \,}}

\title{Sea Contributions to Hadron Electric Polarizabilities through Reweighting}

\ShortTitle{Sea Polarizability via Reweighting}

\author{\speaker{Walter Freeman}\\
        The George Washington University\\
        E-mail: \email{wfreeman@gwu.edu}}

\author{Andrei Alexandru\\
        The George Washington University\\
        E-mail: \email{aalexan@gwu.edu}}

\author{Frank X. Lee\\
	The George Washington University\\
	E-mail: \email{fxlee@gwu.edu}}

\author{Michael Lujan\\
        The George Washington University\\
        E-mail: \email{mlujan@gwu.edu}}

\abstract{
As part of our ongoing lattice study of the electric polarizabilities of hadrons using the background field approach, we use reweighting to examine the effect of the field on the sea quarks. 
As with other reweighting studies, the chief difficulty lies in the construction of a stochastic estimate of the ratio of the fermion determinants. 
In contrast to the case of reweighting in the quark mass, these estimators converge extremely slowly, and are resistant to common variance-reduction techniques such as 
low-mode subtraction. However, it is possible to construct an alternate estimator, taking advantage of the fact that we are interested in only perturbatively small 
fields; this estimator is susceptible to a variance-reduction technique based on a hopping parameter expansion.}

\FullConference{The 30 International Symposium on Lattice Field Theory - Lattice 2012,\\
		June 24-29, 2012\\
		Cairns, Australia}

\begin{document}

\section{Introduction}

Recently there has been a great deal of interest in first-principles lattice computations of hadron polarizabilities.  
The most interesting such quantity is the
electric polarizability of the neutron, since it is difficult to access experimentally due to the lack of free neutrons; the best measurements
have been obtained by neutron-lead~\cite{Schmiedmayer:1991zz} and neutron-deuteron~\cite{Kossert:2002ws} scattering. A basic computation of this quantity
is not too hard to do using the background-field method~\cite{Alexandru:2009id}; the difficulties come in the approach to the physical point that can be compared
with experiment. Effective field theories predict very strong dependence on $m_q$ near the chiral limit, making the neutron polarizability a 
good probe of whether or not chiral behavior is accurately captured by lattice simulations; doing so requires lighter quark masses and potentially
expensive chiral actions~\cite{Lujan:2011ue}. Similarly, we must address finite-volume effects~\cite{Alexandru:2010dx} and finally extrapolate to the continuum. 

The most difficult effect to accommodate, however, is the effect of the electric field on the sea quarks. In principle, this could be done from the
ground up by including the background field (see Sec.~\ref{sec-background}) in gauge generation itself. However, these ensembles will necessarily be uncorrelated.
To determine the polarizability we examine the mass shift in the neutron (or other hadron) when a small background electric field is applied; since the
zero-field and finite-field propagators are strongly correlated, the error on this mass shift can be much smaller than the error on the masses themselves. 
Comparing two uncorrelated ensembles destroys this correlation. What we would like is to generate two (or more) correlated ensembles with different values of $E$.
This can be achieved {\it via} reweighting. 

The most difficult aspect of a reweighting calculation is stochastic estimation of the weight factors; the present work is chiefly concerned with 
this problem.
For reweighting in the quark masses, a straightforward stochastic estimator coupled with several standard variance reduction techniques is generally
successful in obtaining good estimates with a reasonable amount of computer power. These techniques, however, are not useful for reweighting in the
background field. We will discuss possible reasons for those failures, which will likely be illuminating for
other groups performing reweighting computations. We then present an alternative approach in which we construct a stochastic estimator for the derivatives of the weight
factor with respect to the background field which is susceptible to an improvement technique based on a hopping parameter expansion.

The only prior work known to us on sea contributions to polarizability was done with a purely perturbative 
method~\cite{Engelhardt:2010tm}. Instead of applying a 
uniform background field of specified strength throughout the lattice and then computing hadron propagators, the author first expands the path integral
in powers of $E$ and then computes the needed diagrams on the lattice using current insertions. In this method the disconnected and connected insertions
(corresponding to sea and valence contributions) appear quite naturally. The author applies this method in a mixed-action formulation, computing domain-wall
valence propagators on Asqtad dynamical configurations generated by the MILC collaboration~\cite{Bazavov:2009bb} and using stochastic estimators to compute the
disconnected diagrams.


\section{Lattice simulation details}

We apply the following methods to a series of gauge ensembles with two flavors of nHYP-smeared clover 
fermions~\cite{Hasenfratz:2007rf} with $m_\pi \simeq 330$ MeV
and a standard Symanzik-improved gauge action with $\beta = 7.1$, giving $a=0.1255$ fm (determined by the Sommer scale $r_0$)~\cite{Sommer:1993ce}. The ensembles have
volumes $24^3 \times 48$, $32 \times 24^2 \times 48$, and $48 \times 24^2 \times 48$; these elongated lattices were originally generated for a scattering
study~\cite{Pelissier:2012pi}, but we reuse them here, since the elongation in the $x_1$ direction is a convenient probe for finite-volume effects associated with 
Dirichlet boundary conditions. Each ensemble has 300 minimally-autocorrelated configurations.

\section{The background field method}
\label{sec-background}
The most commonly-used approach for computing hadron polarizabilities is the background field method,
allowing the extraction of the polarizability from spectroscopic measurements.
The effect of a uniform background field on the mass of a hadron can be parametrized
as:
\begin{equation}
M_H = -\vec \rho \cdot \vec E -\vec \mu \cdot \vec B - \frac{1}{2} \alpha  E^2 - \frac{1}{2} \beta B^2 + \mathcal O(E^4) + ...
\end{equation}
where $\mu$ is the magnetic moment, $\rho$ the electric dipole moment, $\alpha$ is the electric polarizability, $\beta$ is the magnetic polarizability, and the ellipsis
includes various higher-order terms as well as spin polarizabilities~\cite{Detmold:2009dx,Alexandru:2008sj}. 
While we are mostly interested in the neutron electric polarizability, the reweighting approach 
we will use to probe the effects of the sea is not specific to any particular type of hadron.
The basic approach, then, to extract the polarizability of some hadron is to measure its mass both in the presence and the absence of a
perturbatively small electric field (chosen small enough that higher-order effects are small) and examine the mass shift. 

To apply a background electromagnetic field to the lattice, one can simply apply a U(1) phase on the gauge links on top of the dynamical SU(3) gauge
configurations, making the transformation
\begin{equation}
U_\mu \rightarrow e^{iqaA_\mu} U_\mu.
\end{equation}
We choose here to apply a constant electric field in the $x_1$ direction. This can be done in any suitable gauge; we choose
$A_4 = iEx_1$. The factor of $i$ arises from the Wick rotation to Euclidean time; thus, on the lattice, a real electric field gives
a real factor $e^{\eta x_1}$, while an imaginary field gives pure phase factor. Provided that the magnitude of 
the field is small enough, it should be possible to perform a lattice calculation for an imaginary field and analytically 
continue the results to the real axis; preliminary studies have confirmed that both methods result in consistent results
for the polarizability~\cite{Alexandru:2008sj}. It is convenient to use an imaginary field because the links remain unitary and 
the Dirac operator remains $\gamma_5$-Hermitian.
Thus, we apply the electric field by the transformation
$U_4 \rightarrow e^{-i \eta x_1}$
\noindent where $\eta \equiv a^2 qE$. Note that this value depends on the charge of the quark flavor in question, so to compute the neutron correlator we will need
gauge links for two values of $\eta$.

While in the limit of infinite statistics there is no order-$\eta$ shift in the hadron mass, one may appear from statistical fluctuations in a real
calculation. To eliminate this source of error, we compute hadron correlators for fields $+\eta \hat x_1$ and $-\eta \hat x_1$ (which should, in the 
infinite-statistics ensemble average, be identical) and take their geometric mean to get the two-point function in the presence of the electric field $G(t,\eta)=\sqrt{G_+(t,\eta) G_-(t,\eta)}$.
To extract the polarizability, we then fit it along with the zero-field correlator $G(t,0)$ to the form
\begin{equation}
G(t,\eta) = A e^{-(M_N + \delta \eta^2)t}
\end{equation}
to get the mass shift $\delta$. The two correlators must be fit jointly since they are drawn from the same gauge configurations and their fluctuations are (strongly) correlated; this correlation
greatly reduces the statistical error in $\delta$.

\subsection{Boundary conditions}
 
Periodic boundary conditions are preferable for many lattice observables. However, they potentially create substantial problems
for measurements using background fields. For the gauge we have chosen and for real electric fields, 
there will be an unavoidable and large discontinuity in the 
electric field at the lattice boundary; for imaginary fields, this discontinuity can be avoided only by choosing particular values of $E$~\cite{Rubinstein:1995hc}.
However, such fields are out of the perturbative regime that we want to examine.
For other choices of the gauge, other problems manifest themselves, such as an electric scalar potential that is not single-valued,
leading to manifestly nonphysical scenarios involving quark lines winding around the torus in the direction of the electric field.
It is not clear how one should address these effects. We can avoid them by applying Dirichlet boundary conditions in
the $x_1$ and $x_4$ directions. Dirichlet boundary conditions create their own problems, of course; we can now no longer use certain
improved nucleon sources, such as Coulomb-wall, and can no longer project completely onto a zero-momentum state. However, these effects
can be considered to be finite-size effects which go away in the infinite volume limit, and are manageable in the analysis; we thus 
use Dirichlet boundary conditions~\cite{Alexandru:2008sj}.

\subsection{Choosing a field strength}

\begin{figure}[htb]
\centering\includegraphics[width=0.5\textwidth]{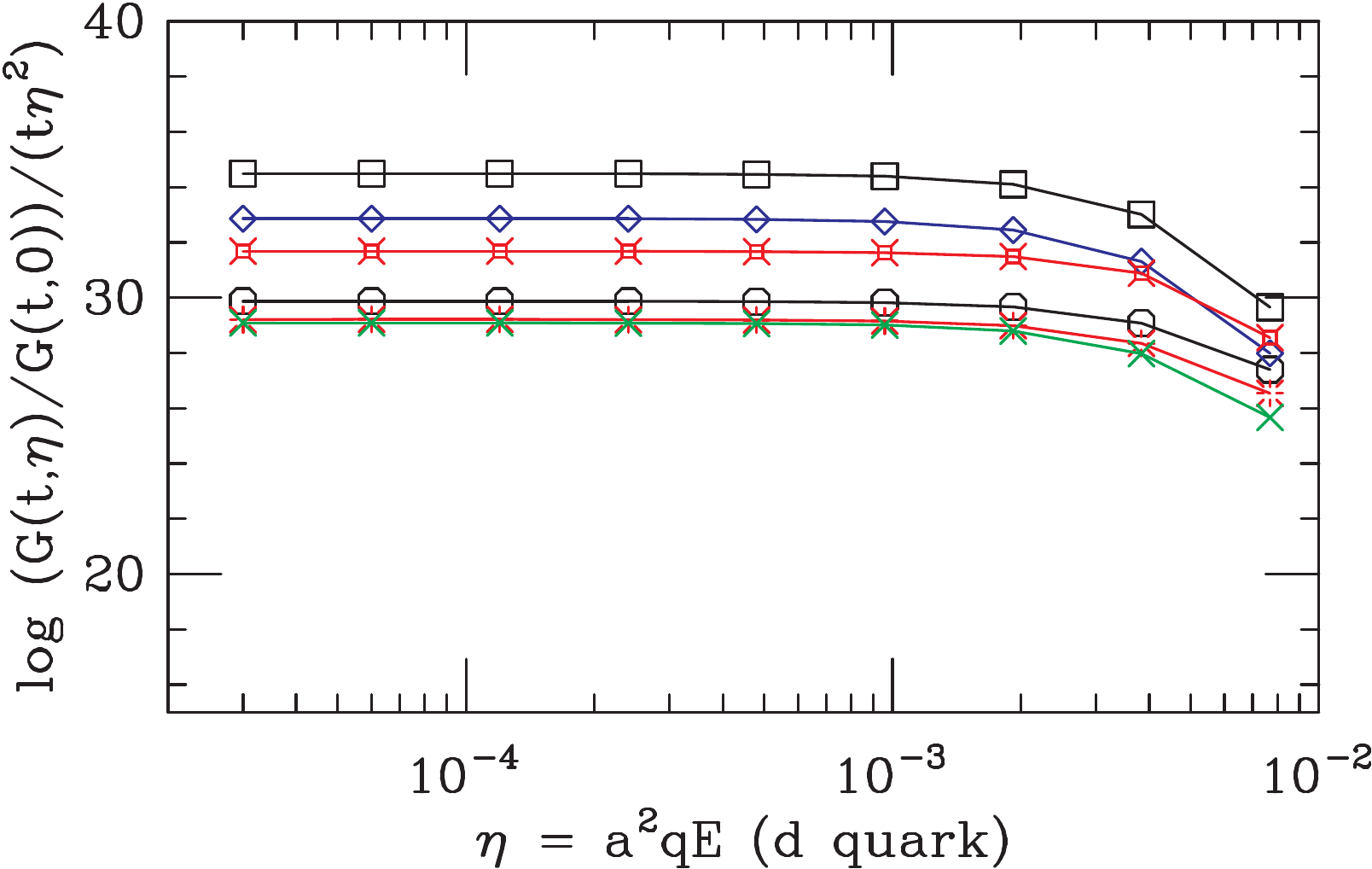}
\caption{Dependence of $(\log \frac{G(t,\eta)}{G(t,0)})/(t \eta^2)$ for the neutral pion on a $24^3 \times 48$ lattice,
as a function of $\eta$ for different correlator times.
This quantity is roughly equivalent to the shift in the 
effective mass divided by $\eta^2$,
and should be constant in the range where $\eta$ creates a purely quadratic effect. The breakdown of 
the quadratic behavior is evident at large $\eta$.}
\label{fig-eta-scaling}
\end{figure}

We must choose a value of the parameter $\eta$ to use for the background field. (Since there are two quark flavors, two $\eta$'s are
required.) Choosing a value which is too large means that we leave the perturbative regime and begin to probe $\mathcal O(E^4)$ effects;
choosing a value which is too small means that we may encounter issues with numerical precision, either with the accuracy of inverters
or (in the extreme case) machine precision. Fig.~\ref{fig-eta-scaling} shows the response of the neutral pion correlator 
to the background field at different temporal separations as a function of the $d$ quark $\eta$ for an inverter precision of $10^{-13}$; the onset of nonperturbative behavior is clear. 
We note that when a similar study was done with an inverter precision of $10^{-9}$, the expected quadratic scaling behavior broke down at
the smallest $\eta$'s. This illustrates the need to choose an $\eta$ small enough to avoid higher-order effects when probing polarizabilities, 
and the need to use a sufficient inverter accuracy to avoid numerical artifacts. Within the large flat region in Fig.~\ref{fig-eta-scaling}, we are free
to choose whatever value of $\eta$ makes the reweighting process perform best.

\section{Reweighting}
Reweighting is a technique for extracting physics based on a different action than the one used in Monte Carlo ensemble generation;
essentially, it allows for {\it post hoc} modification of the parameters in the action. In the standard quantum Monte Carlo,
we would like to do a path integral of the form
\begin{equation}
\LL \mathcal O \RR = \frac{\int[dU] \mathcal O e^{-S_0}}{\int[dU] e^{-S_0}}
\end{equation}
where $S_0$ is the QCD action.

By generating a Monte Carlo ensemble with each configuration weighted by $e^{-S_0}$, the path integral reduces to the familiar 
\begin{equation}
\LL \mathcal O \RR = \frac{\sum \mathcal O}{\sum 1}=\frac{\sum \mathcal O}{N}.
\end{equation}
Suppose, however, that we wish to use this ensemble (weighted by $e^{-S_0}$) to learn about the behavior in the presence
of the background field, given by the action $S_\eta$.
Then the Monte Carlo average gives
\begin{equation}
\LL \mathcal O(S_\eta) \RR = \frac{\sum \mathcal O e^{-(S_\eta-S_0)}}{\sum e^{-(S_\eta-S_0)}}
\end{equation}
where $e^{-(S_\eta-S_0)}$ is a ``weight factor'' which may be interpreted as the relative prominence of a particular configuration
in the target and source distribution. If there is little overlap between the source and target ensembles, reweighting will fail
as the weight factors fluctuate wildly (over many orders of magnitude); in general, reweighting always comes with a decrease in 
statistical power, as described in~\cite{Liu:2012gm}. This may not always be immediately evident and it is
possible to wind up underestimating statistical errors~\cite{Hasenfratz:2008fg}. Reweighting is used to reweight in the quark mass 
to approach the chiral limit without incurring the expense of simulating at those light quark masses 
directly~\cite{Hasenfratz:2008fg}, 
to perform simulations at small but nonzero chemical potential, 
to couple sea quarks to dynamical photon fields to probe
isospin breaking~\cite{Ishikawa:2012ix}, and (most similarly to this project) been used to investigate the intrinsic strangeness of the nucleon~\cite{Ohki:2009mt}.  In this
last case, the authors sought to evaluate $\PAR{M_N}{m_s}$ by measuring $M_N$ at slightly different values of $m_s$ and examining the
difference. This is difficult if the errors in $M_N$ are uncorrelated, but by reweighting in the strange quark mass to generate correlated
ensembles they were able to measure errors on the difference in $M_N$ more accurately than $M_N$ itself. We intend to do a very similar
thing, except with $\eta$ (which can be thought of as either reweighting in the background field or the quark electric charge) 
instead of in the strange quark mass.

\subsection{Estimating the weight factor}

The weight factor $w_i$ is given by $e^{-(S_\eta-S_0)_i} = \det \frac{M_\eta}{M_0} = \det^{-1} M_0 M_\eta^{-1}$, where $M_0$ and $M_\eta$
are the fermion matrices corresponding to the actions $S_0$ and $S_\eta$. This determinant is impractical to compute exactly
and must be estimated stochastically. The standard stochastic estimator for the inverse determinant of some matrix $\Omega$ is~\cite{Hasenfratz:2008fg}
\begin{equation}
(\det \Omega)^{-1} = \LL e^{-\xi^\dagger(\Omega-1)\xi} \RR.
\end{equation}

While the inverse determinant of $M_0 M_\eta^{-1}$ is real, using this operator with the above estimator will produce complex results. We thus instead
apply the stochastic estimator to an operator with the same determinant but which is Hermitian: $\Omega = \sqrt{M_\eta^{\dagger-1} M_0^\dagger M_0 M_\eta^{-1}}$.
The square root can be taken by a rational function approximation as done in rational hybrid Monte Carlo, etc. This is somewhat computationally expensive,
as it requires inverting a matrix which itself requires the computation of inversions; however, as that matrix is in principle perturbatively close to 1 its
inversion should not require too many iterations. As an alternative, the authors of~\cite{Finkenrath:2012cz} have argued that 
in such cases one should simply discard the imaginary part of the estimator, saving all the work associated with the rational function approximation. 
In the analysis
that follows, we estimate the inverse determinant of $\Omega = \sqrt{M_\eta^{\dagger-1} M_0^\dagger M_0 M_\eta^{-1}}$ where $w_i = \det^{-1} \Omega$.

While this stochastic estimator may be (and generally is) quite noisy, this noise does not introduce bias because the average over the noise vectors $\xi$
commutes with the gauge average~\cite{Hasenfratz:2008fg}; in principle, if the ensemble is large enough, one need only use a single stochastic estimate per configuration,
although with a gauge ensemble of finite size it is often profitable to work harder than this to improve the stochastic estimator to reduce its fluctuations.
The simplest way to do this is to average multiple stochastic estimates, but there are other techniques that can result in a greater reduction in the stochastic
noise for a given budget of computational power.
We note that the ``signal'' we are trying
to extract from these estimates is the true fluctuation of the weight factor from one configuration to the next, and thus we may adopt the rough criterion 
for reduction of the stochastic noise that
$\sigma_{\rm{gauge}} \gtrsim \sigma_{\rm{noise}}$. This condition is not necessary, but it should be sufficient, for a successful application of reweighting.
(Note, however, that if the true values of the weight factor are indeed very similar, it is possible for this test to give an overly-pessimistic description
of the quality of the stochastic estimates.)

\section{Improving the stochastic estimate}

\begin{figure}
\centering\includegraphics[width=0.3\textwidth]{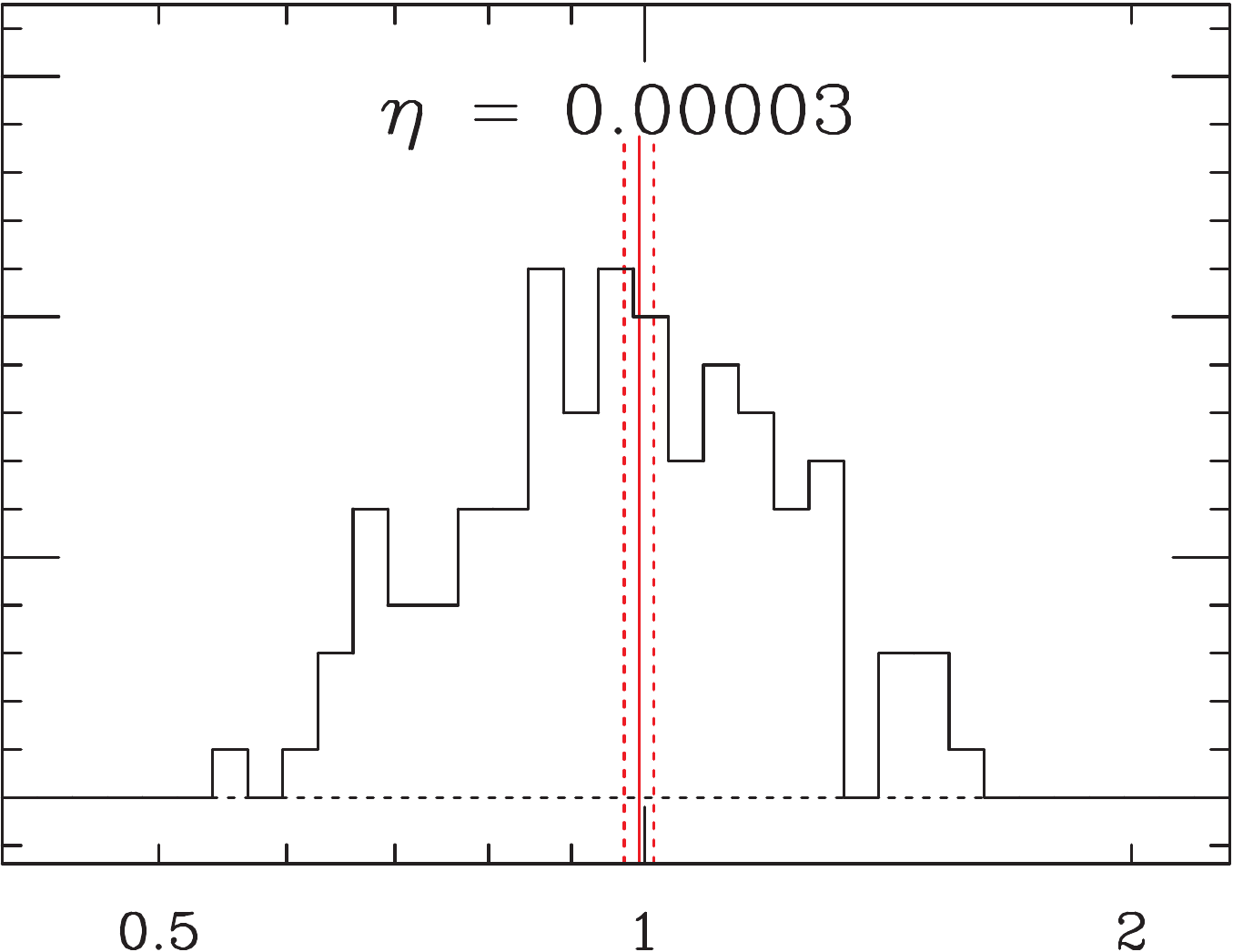}
\includegraphics[width=0.3\textwidth]{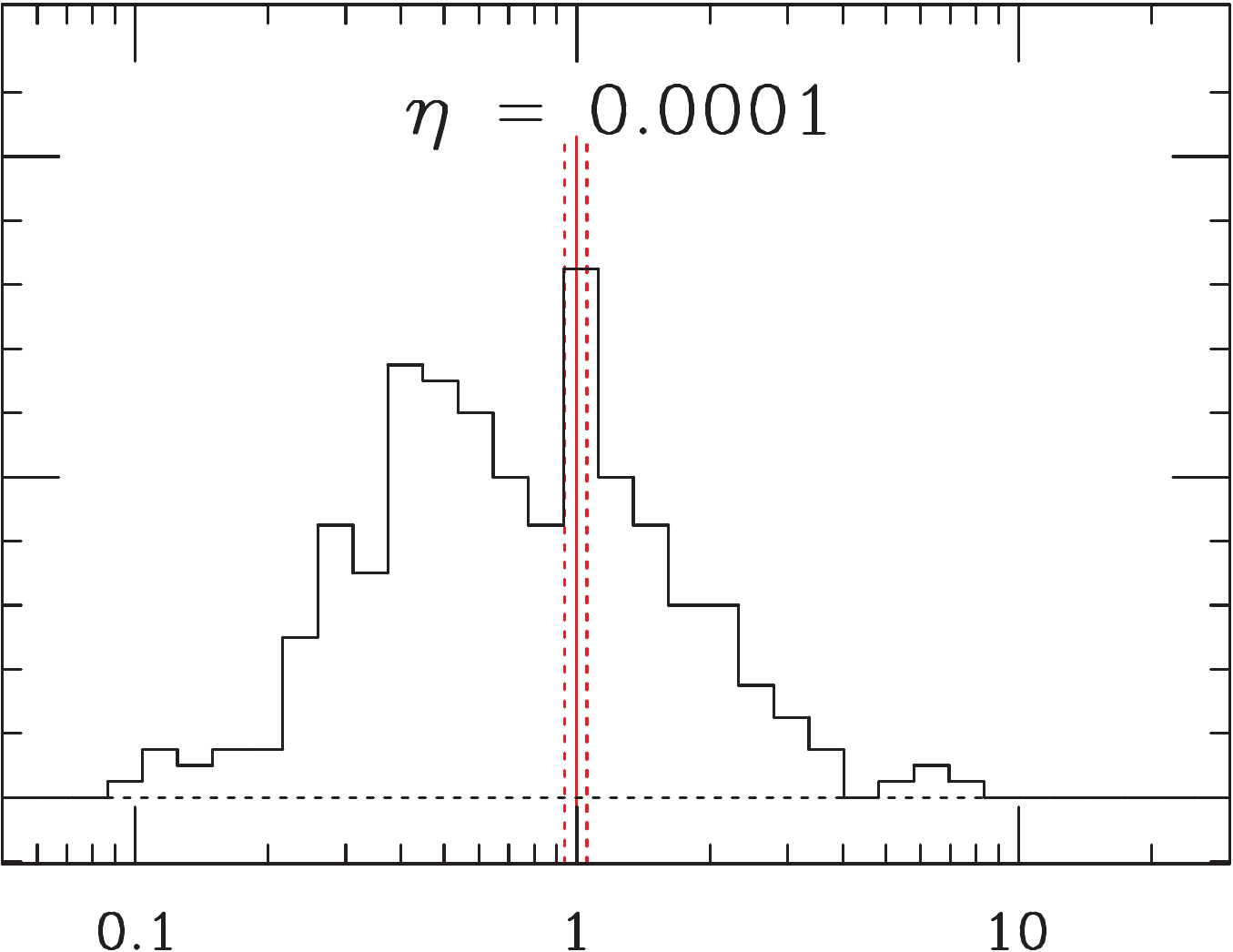}
\includegraphics[width=0.3\textwidth]{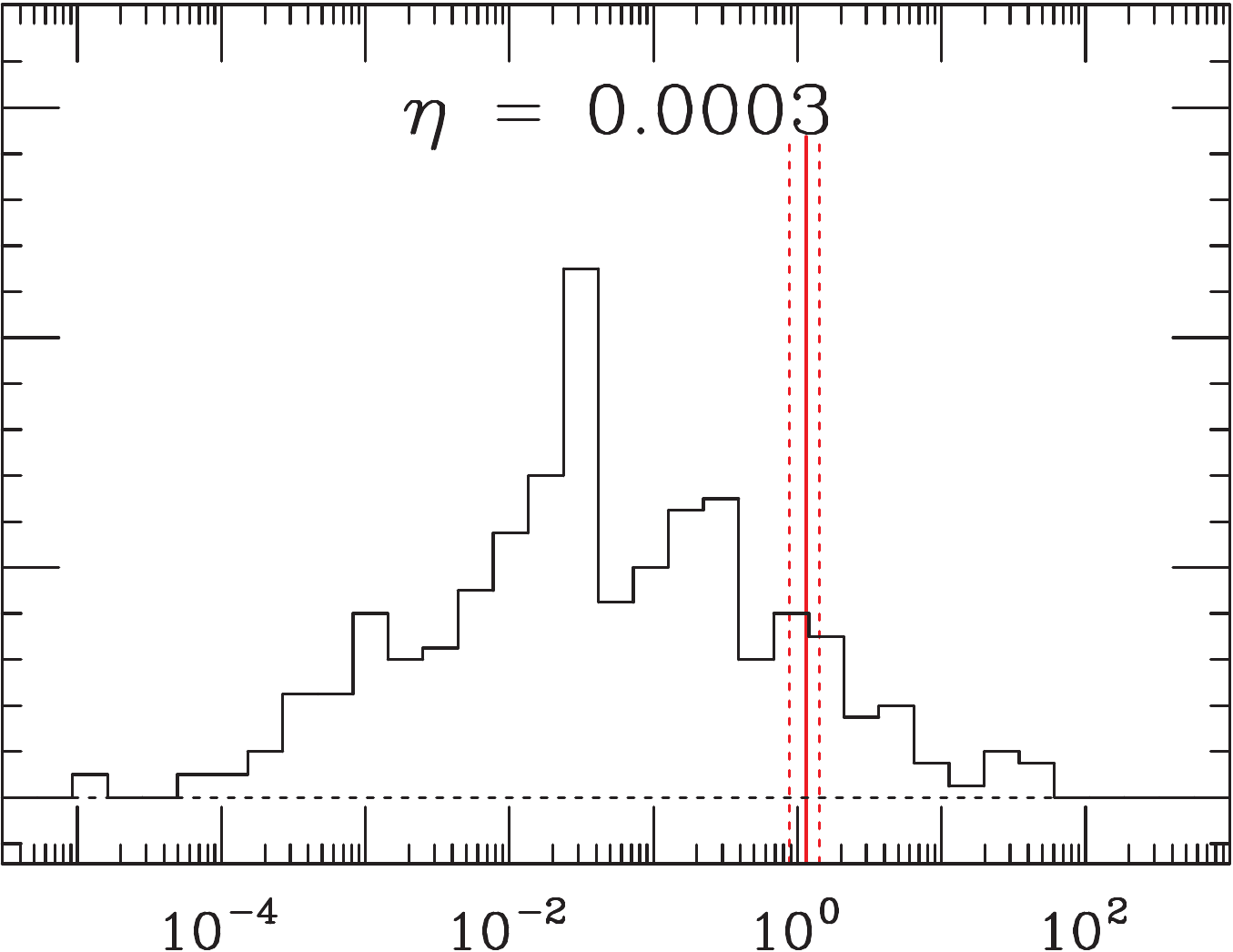}
\caption{Histograms of the stochastic estimator of the weight factor 
$e^{-\xi^\dagger(\Omega-1)\xi}$
for three different values of $\eta$, along with
the mean and its standard error, on a single $24^3 \times 48$ configuration. Even with a large number of noises, the average is indistinguishable from unity; for
large values of $\eta$, the average is dominated by the few points in the right tail of the distribution, which is roughly log-normal.}
\label{fig-fluct}
\end{figure}

There is a fundamental problem, however: this estimator is tremendously noisy in our case. As one might expect, the distribution of stochastic estimates $e^{-\xi^\dagger(\Omega-1)\xi}$
is log-normal; when the width is large enough that the skew of the distribution is apparent, sampling the long tail becomes very difficult. For small values of 
$\eta$, the average estimate of the weight factor is very close to unity, while for larger values the long-tail sampling problem becomes tremendously difficult; see Fig.~\ref{fig-fluct}.\footnote{While it is of course possible that the true value of the weight factor on the configuration shown here is close to unity, tests on other configurations give the same result.}
We have confirmed that the estimator does produce the correct result for the determinant on a $4^4$ lattice where that determinant can be computed exactly~\cite{Alexandru:2010yb}, but it required $10^5$-$10^6$ noise
vectors, something clearly unaffordable for production-size lattices!

\subsection{Low-mode subtraction}

A commonly-used technique in reweighting in the quark mass is low-mode subtraction. By projecting out the low modes of the operator in question and computing their inverse determinant exactly, 
we stand to improve the stochastic estimator in two ways. First, we accelerate the inversions required to make the estimates by 
reducing the conditioning number of the operator being inverted, allowing for the computation of a larger number of estimates with
the same amount of computer time. Second, if an appreciable fraction of the fluctuations of the estimator come from the low modes,
we may eliminate those fluctuations by treating the low sector exactly (by simply multiplying together the eigenvalues). 
This is offset, of course, by the additional cost of computing the eigensystem. Can this technique be profitably applied to 
reweighting in the sea quark charges?

Computation of the eigensystem is more difficult in our case. In the mass-reweighting case, the eigenvectors for all values of $\kappa$ are the same,
so the eigenvectors of $\Omega$ are the same as the eigenvectors of the Dirac operator. On the other hand, we must do a separate calculation to compute the eigensystem of 
$\Omega = \sqrt{M_\eta^{\dagger-1} M_0^\dagger M_0 M_\eta^{-1}}$, which is quite computationally expensive. Furthermore, the eigenvectors
differ for different values of $\eta$, causing problems with the determinant breakup method (see Sec.~\ref{sec-breakup}). 

\begin{figure}[htb]
\centering
\vspace{0pt}
\begin{minipage}[t]{0.42\textwidth}
\vspace{0pt}
\includegraphics[width=1\textwidth]{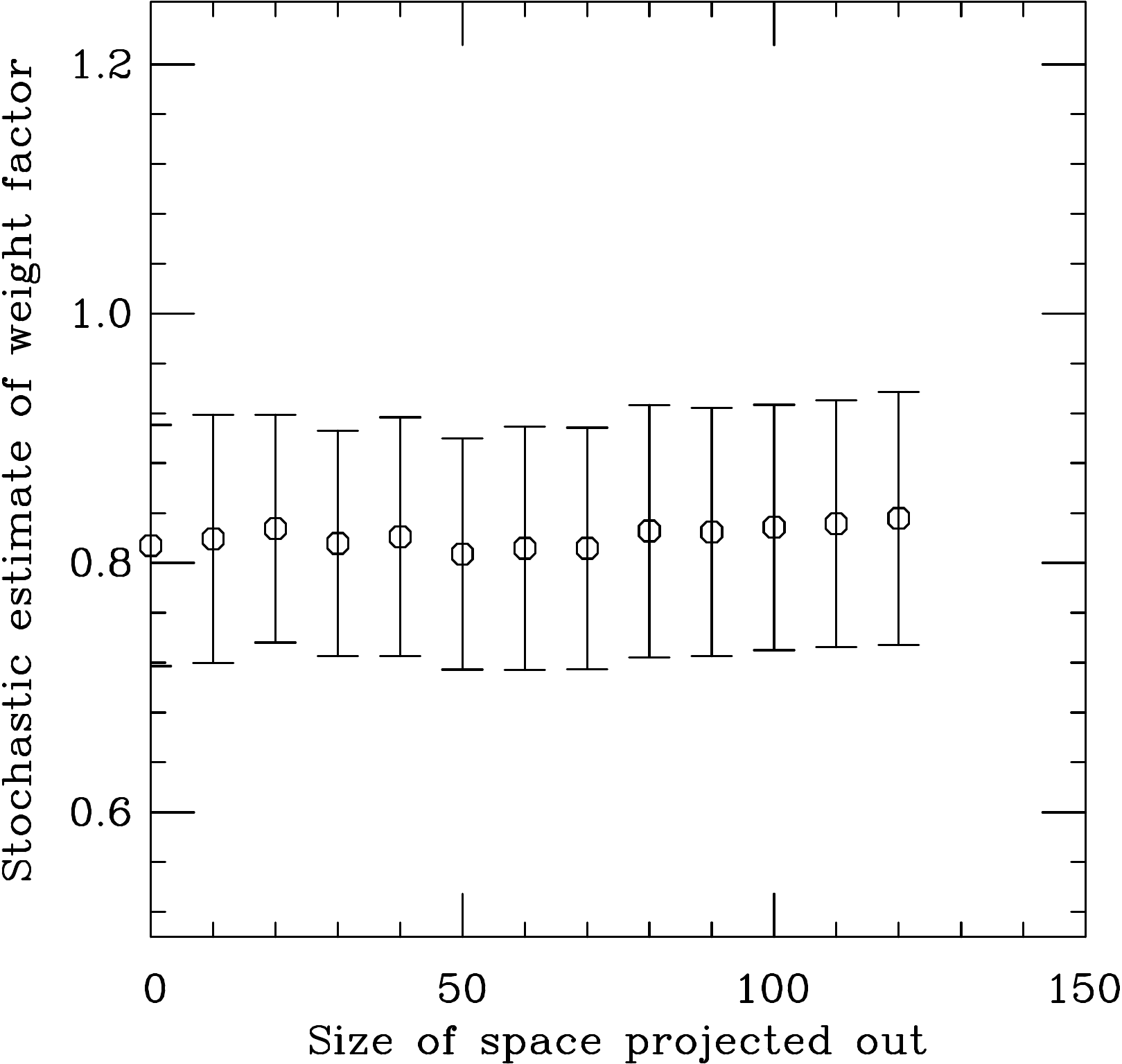}
\end{minipage}
\begin{minipage}[t]{0.42\textwidth}
\vspace{0pt}
\includegraphics[width=0.9\textwidth]{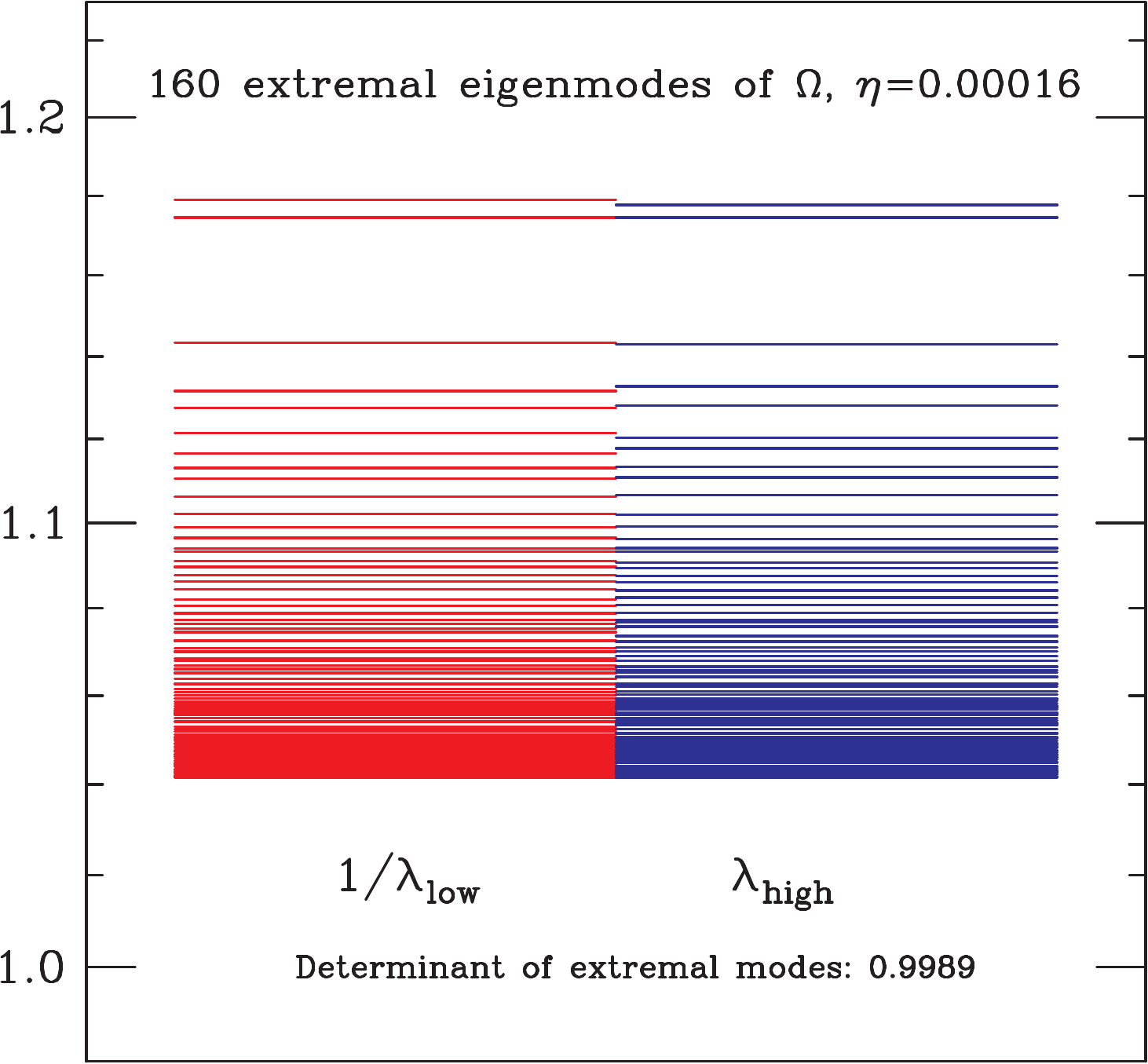}
\end{minipage}
\caption{Left: Dependence of the stochastic estimator of the weight factor on the number of extremal modes projected out, showing no meaningful decrease in the variance, on a representative
configuration and value of $\eta$. The same noise vectors were used for each point. 
Right: Highest and lowest eigenvalues of $\Omega \equiv \sqrt{M_\eta^{\dagger-1} M_0^\dagger M_0 M_\eta^{-1}}$ for $\eta=0.00018$, showing pairing between high
and low modes.
}
\label{fig-spectrum}
\label{fig-proj}

\end{figure}

Since our matrix $\Omega$ is in principle close to the identity, we computed eigensystems for both high and low sectors to test the method,
generated an ensemble of a few hundred noise vectors, and used those to estimate $\det^{-1} \Omega$ varying the size of the extremal sectors
treated exactly. 
Specifically, if $P$ is a projector onto the space spanned by $N$ extremal modes of $\Omega$ with eigenvalues $\lambda_i$, and $\overline P \equiv 1-P$, we may write
\begin{equation}
(\det \Omega)^{-1} 
= \left( \prod_{i=1}^N \frac{1}{\lambda_i} \right) \LL e^{-\xi^\dagger \overline P (\Omega-1) \overline P \xi} \RR.
\end{equation}
To our surprise, there is essentially no effect whatsoever on the size of the stochastic fluctuations, as shown in Fig.~\ref{fig-proj}.
This is strikingly different from the behavior for mass reweighting~\cite{Hasenfratz:2008fg}.

In mass reweighting, most of the signal comes from the low modes, as shown empirically in~\cite{Ohki:2009mt}; in fact, almost the entirety of their signal comes
from the low modes, and while the high modes are included for correctness their contribution seems to be only (a small amount of) noise. The authors of~\cite{Luscher:2008tw}
show this analytically, and furthermore argue that in the case of mass reweighting most of the noise comes from the low modes as well; 
the estimator is protected from large fluctuations from the high modes. Thus, by removing the low modes from the stochastic estimator, it is possible to also
remove most of the noise. This is only guaranteed in the case of reweighting in the quark mass, however; in our case, it fails. Since changing the quark mass does not 
change the eigenvectors, the low modes of the Dirac operator itself are also eigenvectors of the matrix $\Omega$ for quark mass reweighting; this is not true for us,
and the extremal modes of $\Omega$ may not be related to the low modes of the Dirac operator.

For reweighting in the 
sea quark charge, the extremal modes of $\Omega$ do not contribute much to the determinant. This is because they are very nearly paired; for each high mode with eigenvalue $\lambda$,
there is a low mode with an eigenvalue close to $1/\lambda$, nearly cancelling their contribution to the determinant; a representative case is shown in Fig.~\ref{fig-spectrum}. 
Studies on $4^4$ and $6^4$ lattices confirm that most of the signal (the difference of the determinant from unity) in our case comes from the bulk ``interior'' modes of $\Omega$.


\subsection{Determinant breakup}
\label{sec-breakup}
Another commonly-used technique for improving the stochastic estimator is to divide the reweighting up into many steps and compute an independent estimate of each one;
this technique is often referred to as ``determinant breakup''~\cite{Hasenfratz:2008fg} or 
``determinant factorization''~\cite{Liu:2012gm,Alexandru:2002jr}. 
For reweighting in the quark charge, for instance, one might imagine estimating the 
weight factor to reweight from $\eta=0$ to $\eta=\eta_1$, then the weight factor from $\eta=\eta_1$ to $\eta=\eta_2$, and so forth until the desired value is reached.
However, for reweighting in $\eta$, this technique also fails to decrease the stochastic noise in the estimator compared to simply using more noise vectors.
The authors of~\cite{Liu:2012gm} find that for reweighting in a substantial shift in $m_l$, it is more efficient to use more subintervals in the determinant breakup (which they call ``steps``) than 
repetitions of the entire procedure (``hits'') for a given total number of inversions. However,
this improvement is limited, as shown in~\cite{Finkenrath:2012cz}, 
and, as predicted in~\cite{Liu:2012gm}, there is no more to be gained after $N_{\rm {steps}}$ increases past a certain point. We suspect that this behavior occurs
because determinant breakup obtains its benefits by converting a large reweighting step into a sequence of small ones; once the
steps are already sufficiently small, no further benefit is gained by splitting them up further compared to simply increasing $N_{\rm {hits}}$. 
There is no benefit to applying this technique to reweighting in $\eta$, however, and this explains why; we are already reweighting by a perturbatively 
small interval in order to make a valid measurement of the polarizability.

\section{The perturbative estimator}

Since it is not feasible to perform a direct stochastic estimation of the determinant, and since neither of the most common improvement techniques that
are successful for mass reweighting work for reweighting in the quark charges, we turn to an alternative estimator. We are interested only in the weight
factors for perturbatively small values of $\eta$, so we can expand about $\eta=0$:
\begin{equation}
w(\eta) = 1 + \eta \left. \PAR{w}{\eta} \right|_{\eta=0} + \frac{1}{2} \eta^2 \left. \PARTWO{w}{\eta} \right|_{\eta=0} + ...
\end{equation}
The expansion must be taken to second order in $\eta$, since we are interested in effects of order $E^2$. The task then becomes
to construct estimators for $\left. \PAR{w}{\eta} \right|_{\eta=0}$ and $\left. \PARTWO{w}{\eta} \right|_{\eta=0}$. Both terms are needed,
since the linear term in the weight factor can couple with a linear dependence on $\eta$ to give a quadratic effect, or the quadratic 
term can give a second-order effect on its own. Given estimates of these derivatives, we can evaluate the above at any sufficiently-small
$\eta$ to produce a reweighted ensemble on which to apply the valence calculation. 

To construct such an estimator for $\left. \PAR{w}{\eta} \right|_{\eta=0} = \left. \PAR{}{\eta} \frac{\det M_\eta}{\det M_0}\right|_{\eta=0}$,
we rewrite $\det M_\eta$ as a Grassmann integral:
\begin{align}
\PAR{}{\eta} \det M &= \PAR{}{\eta} \int d\psi d\bar \psi \, e^{-\bar \psi M \psi} \nonumber \\
&= \int d\psi d\bar \psi \, \left( -\bar \psi \PAR{M}{\eta} \psi \right) e^{-\bar \psi M \psi} \nonumber \\
&=\det M \, {\mathrm  {Tr}} \left(\PAR{M}{\eta} M^{-1} \right).
\end{align}
This is an expression for $\PAR{\det M}{\eta}$, but we want $\left. \frac{1}{\det M_0} \PAR{}{\eta} {\det M}\right|_{\eta=0}$. When the above is evaluated
at $\eta=0$, the determinant in front cancels the $\frac{1}{\det M_0}$ and we get 
\begin{equation}
\left. \PAR{}{\eta} \frac{\det M_\eta}{\det M_0}\right|_{\eta=0} = {\mathrm  {Tr}} \left( \left. \PAR{M}{\eta}\right|_{\eta=0} M_0^{-1} \right)
\end{equation}
The trace involved here must still be evaluated stochastically. 

The second derivative term proceeds similarly. Computing $\PARTWO{}{\eta} \det M$, we get
\begin{align}
\PARTWO{}{\eta} \det M =& \PARTWO{}{\eta} \int d\psi d\bar \psi \, e^{-\bar \psi M \psi} \nonumber \\
=& \PAR{}{\eta} \int d\psi d\bar \psi \, \left( -\bar \psi \PAR{M}{\eta} \psi \right)\, e^{-\bar \psi M \psi} \nonumber \\
=&              \int d\psi d\bar \psi \, \left( -\bar \psi \PARTWO{M}{\eta} \psi \right) e^{-\bar \psi M \psi} \label {term1}\\
&+ \int d\psi d\bar \psi \, \left( \bar \psi \PAR{M}{\eta} \psi \right)^2 e^{-\bar \psi M \psi} \label{term2}
\end{align}
Evaluating these integrals gives:
\begin{equation}
\PARTWO{}{\eta} \det M = \det M \left[ \Tr \left(\PARTWO{M}{\eta} M^{-1}\right) - \left( \Tr \PAR{M}{\eta} M^{-1}\right)^2 + \Tr \left(\PAR{M}{\eta} M^{-1} \right)^2\right].
\end{equation}
As above, this gives
\begin{equation}
\left. \PARTWO{}{\eta} \frac{\det M}{\det M_0}\right|_{\eta=0} = \Tr (BM_0^{-1}) + \Tr (AM_0^{-1} AM_0^{-1}) - \left( \Tr (AM_0^{-1}) \right)^2
\end{equation}
where $A \equiv \left. \PAR{M}{\eta} \right|_{\eta=0}$ and $B \equiv \left. \PARTWO{M}{\eta} \right|_{\eta=0}.$ 

\begin{figure}
\centering\includegraphics[width=0.5\textwidth]{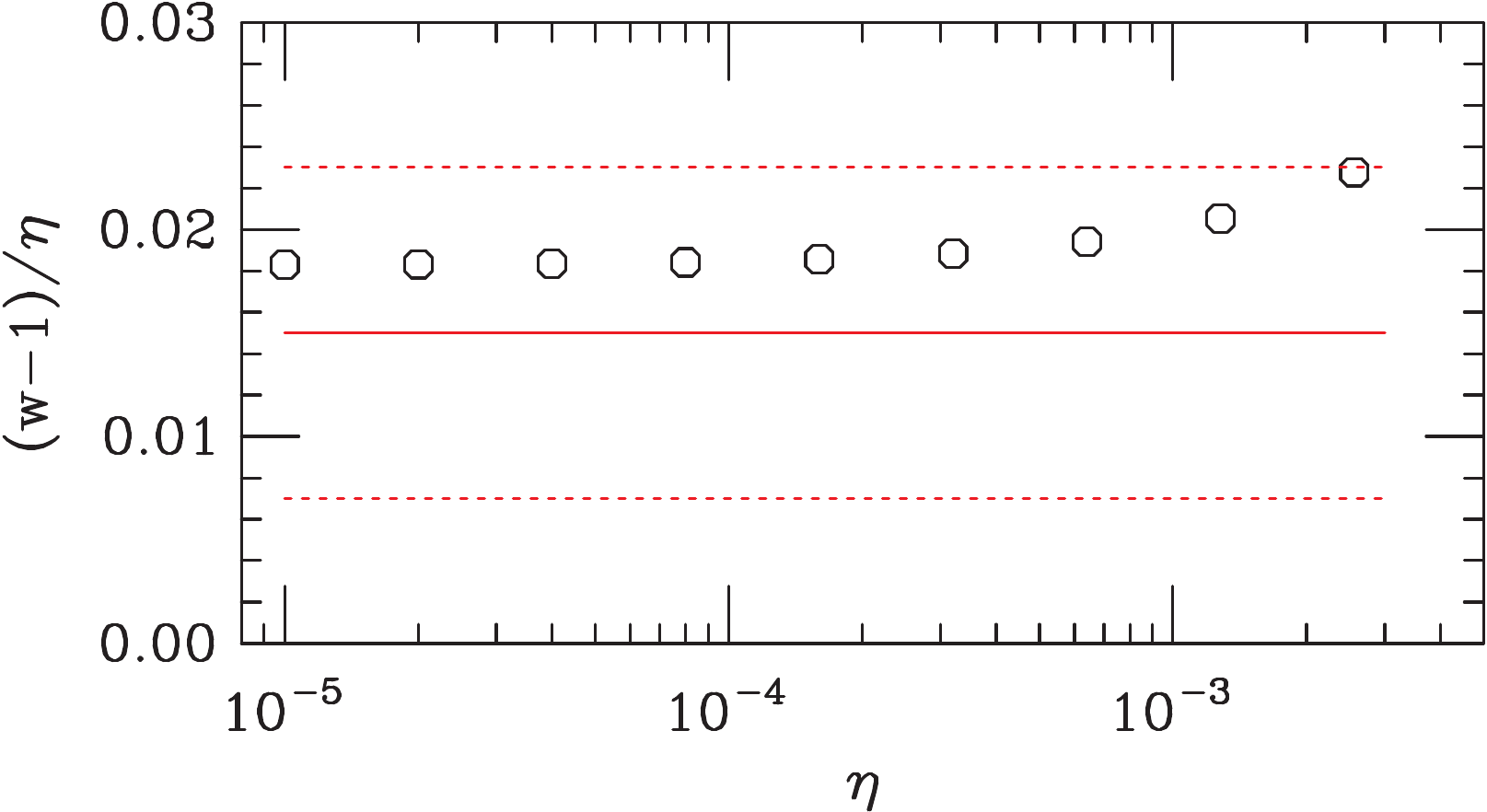}
\caption{Exact values for $\frac{\det M_\eta/\det M_0-1}{\eta}$ on a $4^4$ lattice, compared with the value predicted by the stochastic estimator for $\PAR{\det M_\eta/\det M_0}{\eta}$
and its error band.}
\label{deriv-confirm}
\end{figure}

Now we must construct stochastic estimators for these three traces, as well as the trace in the first derivative. We use the standard stochastic 
estimator $\Tr\,\mathcal O = \LL \xi^\dagger \mathcal O \xi \RR$; note that two stochastic estimates of the first derivative term can be used to 
construct an estimate of $\left( \Tr (AM^{-1}) \right)^2$. Tests on a $4^4$ lattice confirm that this estimator for the first derivative
is correct; see Fig.~\ref{deriv-confirm}. However, it is just as noisy as the direct estimator of the determinant; the data in Fig.~\ref{deriv-confirm} required $5 \times 10^6$ 
noise vectors to produce! Clearly this new estimator on its own has not gained anything.

\subsection{Hopping parameter expansion improvement}

However, this estimator for the trace is susceptible to an improvement technique. 
If other operators $\mathcal O'_i$ can be identified such that the stochastic fluctuations in $\xi^\dagger \mathcal O \xi$ and $\xi^\dagger \mathcal O'_i \xi$ are correlated,
then we can reduce the overall fluctuations by writing
\begin{equation}
\Tr \mathcal O = \LL \xi^\dagger \left(\mathcal O - \sum_i \mathcal O'_i \right)\xi \RR + \sum_i \Tr \mathcal O'_i.
\label{imp-def}
\end{equation}
Obviously, for this to be useful, the $\mathcal O'_i$'s must themselves have exactly-computable traces.

For the operators needed here, a set of improvement operators can be gotten by performing a hopping parameter expansion of $M^{-1}$, with
each term in the expansion acting as one $\mathcal O'_i$; this guarantees that the fluctuations in the estimator will be correlated. Specifically, the 
fluctuations are correlated with the magnitude of the off-diagonal elements of $\PAR{M}{\eta} M^{-1}$; the terms in the hopping parameter expansion 
approximate the largest of these close to the diagonal~\cite{Thron:1997iy}. For the
Wilson-clover fermions considered here, we may write
$M$ as $1 - \kappa \left( \mathcal D + C \right)$, where the hopping term $\mathcal D$ is given~by
\begin{equation}
\mathcal D_{mn} = \sum_\mu \left[ (1-\gamma_\mu) U_\mu(m) \delta_{m + \hat \mu, n} +  (1+\gamma_\mu) U_\mu^\dagger(n) \delta_{m,n+\hat \mu} \right] 
\end{equation}
and the clover term $C$ is given by
\begin{equation}
C_{nn} = c_{SW} \frac{1}{8} \sum_{\mu \nu} \sigma_{\mu \nu} L_{\mu \nu}(n)\,,
\label{def-clover}
\end{equation}
where $L_{\mu \nu}(n)$ is a sum of the imaginary part of the plaquettes in the $\mu \nu$ plane that include site $n$.

For the first derivative term, the technique of~\ref{imp-def} gives
\begin{align}
\Tr \PAR{M}{\eta} M^{-1} =& \LL \xi^\dagger \left( \PAR{M}{\eta} M^{-1} - \PAR{M}{\eta} - \kappa \PAR{M}{\eta} (\mathcal D+C)
- \kappa^3 \PAR{M}{\eta} (\mathcal D+C)^2 - \kappa^2 \PAR{M}{\eta} (\mathcal D+C)^3 - ... \right) \xi \RR \nonumber \\
&+ \Tr \PAR{M}{\eta} + \Tr \kappa \PAR{M}{\eta} (\mathcal D+C) + \Tr \kappa^2 \PAR{M}{\eta} (\mathcal D+C)^2 + \Tr \kappa^3 \PAR{M}{\eta} (\mathcal D+C)^3 + ...
\end{align}
where $\mathcal O'_i = \PAR{M}{\eta} M^{-1} \kappa^i (\mathcal D+C)^i$.
The stochastic component of this can be computed directly as before. 
The difficult part is to compute $\Tr \kappa^n \PAR{M}{\eta} (\mathcal D+C)^n$ quickly. 
In the end we wish to apply this technique to nHYP-smeared Wilson-clover fermions, but it is useful
to consider the unimproved Wilson action first, for which $C=0$. 

The trace of $\PAR{M}{\eta} \mathcal D^n$ can be evaluated most simply by writing $\mathcal D$ as the sum of eight hopping terms. $\PAR{M}{\eta}$
can be calculated analytically, and consists itself of two hopping terms in the $\pm \hat x_4$ direction; the rest are zero. We break $\PAR{M}{\eta}$ and $\mathcal D^n$
into separate hopping terms and expand the product. All terms which give a nonzero contribution to the trace are products of hops which form closed paths.
However, not all such paths contribute; some closed paths (such as those that double back on themselves) have zero Dirac trace and do not contribute. The most
efficient way to compute the trace is thus to expand $\PAR{M}{\eta} \mathcal D^n$ as a sum of products of hopping terms which have definite spatial and Dirac structure,
each corresponding to the product of gauge links along a path of a particular shape,
and pick out only those paths which are closed. We then compute the Dirac trace of each term. The computationally-intensive step, summing the appropriate products of 
gauge links over the spatial volume, must only be done for terms with nonzero Dirac trace.

The clover term can be dealt with by considering a generalization of the procedure used for the pure Wilson action. Another way to describe that procedure is as a separation of the $\mathcal D$ operator
into eight pieces (hops in $\pm x,y,z,t$), each with a definite Dirac and spatial structure. The need to separate out terms with a definite spatial structure is readily apparent since in taking the trace we are only
concerned with closed paths. In doing so we got a definite Dirac structure that does not depend on the gauge links and can be factored out of the sum over sites along the way.
The same principle applies to the clover term even though the spatial structure is trivial: separate $C$ into six pieces each with definite Dirac structure. These are the individual terms $\sigma_{\mu \nu} L_{\mu \nu}$ in 
Eq.~\ref{def-clover}, allowing us to factorize the trace into Dirac and SU(3) pieces. 

We must also compute $\PAR{L_{\mu \nu}}{\eta}$, as it appears in $\PAR{M}{\eta}$. This can be done analytically for ``plain'' clover fermions, but
for nHYP-smeared clover fermions, the smearing process complicates computation of $\PAR{\mathcal D}{\eta}$ and $\PAR{C}{\eta}$. While this can in principle
still be done analytically, it is simpler and not expensive to do 
it
numerically.
Our procedure for evaluating $\Tr \kappa^n \PAR{M}{\eta} (\mathcal D+C)^n$ is thus as follows:

\begin{enumerate}
\item{Write $\mathcal D$ and $\PAR{\mathcal D}{\eta}$ (the latter calculated numerically) as the sum of eight separate hopping terms.
Similarly, break $C$ and $\PAR{C}{\eta}$ into six terms, each with the Dirac structure of $\sigma_{\mu \nu}$.}
\item{Expand $\PAR{(\mathcal D+C)}{\eta}(\mathcal D+C)^n$, giving $14^{n+1}$ terms, each with a definite Dirac and spatial structure.}
\item{Compute spatial part: The majority of these terms correspond to paths that are not closed, and can be discarded right away.}
\item{Each term now can be factorized into a SU(3) part (which depends on the gauge links, and must be summed over sites), and a Dirac part (which is the same for each lattice site and can be factored out).}
\item{Compute Dirac part: the majority of terms will have zero Dirac trace and can be discarded.}
\item{Compute SU(3) part: For the remaining terms (roughly one in 500, regardless of order, after the first few), do the hard work of computing products of links and $L_{\mu \nu}$'s over all sites.}
\end{enumerate}

\noindent The second-derivative traces can be evaluated in a very similar way. For the $\Tr (BM^{-1})$ term, we must additionally compute $B \equiv
\PARTWO{(\mathcal D+C)}{\eta}$, but this is not difficult to do numerically. 

\begin{figure}
\centering\includegraphics[width=0.55\textwidth]{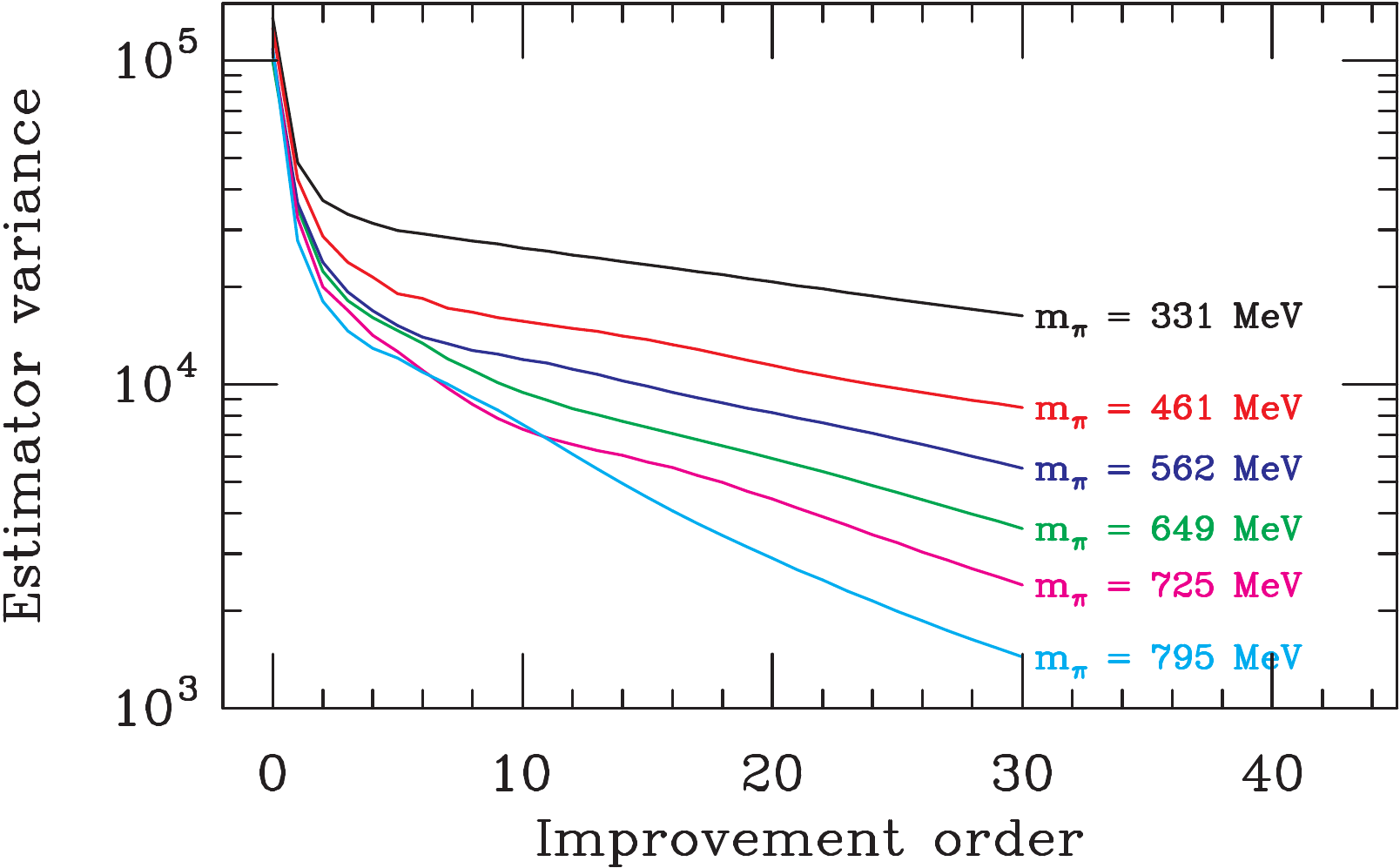}
\caption{
Improvement in the variance of the stochastic estimator of $\PAR{w}{\eta}$ as a function of the order of the hopping parameter expansion for different
values of $m_\pi$.}
\label{fig-scaling}
\end{figure}

This expansion can be in principle carried out to arbitrarily high order. The evaluation of the stochastic estimates of additional $\mathcal O'_i$'s is 
not difficult, as the cost is dominated by the inversion which only must be performed once for any number of orders. 
The limiting factor is the evaluation of $\Tr \kappa^i \PAR{M}{\eta} \mathcal D^i$, as the cost increases exponentially with $i$. 
However, it is possible to examine the reduction in stochastic noise without computing the exact traces. We find that the degree to which
this improvement procedure reduces the stochastic noise, especially after the first few orders, is strongly dependent on the value of $m_\pi$;
this is not surprising in light of the use of a hopping parameter expansion. This behavior is shown for several different values of $m_\pi$ in Fig.~\ref{fig-scaling}. 
Because the improvement slows down after the first few orders and the cost increases substantially, we compute the exact traces only up to seventh order.

\section{Applicability to the polarizability and future plans}
While the improvement technique carried out to seventh order yields a substantial reduction in the variance of the stochastic estimator,
this estimator is still quite expensive to run; we aim for around $10^3$ stochastic sources per configuration. 
These estimates are almost complete on the $24^3 \times 48$ ensemble. Once they are complete, we will use the estimates of $\PAR{w}{\eta}$ and $\PARTWO{w}{\eta}$
to evaluate $w(\eta)=\frac{\det M_\eta}{\det M_0}$ at particular values of $\eta$ (corresponding to the charges of the up and down sea quarks), 
chosen small enough that they lie within the perturbative regime of Fig.~\ref{fig-eta-scaling}
but large enough to avoid any subsequent numerical precision issues, and then use the resulting weight factors to generate a reweighted 
ensemble in the presence of the background field. The remainder of the polarizability computation proceeds as in 
Ref.~\cite{Alexandru:2009id},
but using the reweighted ensemble to compute the nonzero-field neutron correlator. 
It is likely that the influence of the charged sea is greater at lighter quark masses. We have such an ensemble available (with $m_\pi \approx 200$ MeV) and 
intend to repeat the calculation on it, although the variance reduction from the hopping-parameter expansion will not be as strong due to the lower $m_\pi$.
Finally, once the calculation detailed above is complete, if the variance in the stochastic estimator of the weight factors leads to a large increase in the
overall error, we are considering using deflation to accelerate the thousands of inversions required to further improve the estimator.

\section{Acknowledgements}

We would like to acknowledge Craig Pelissier for his assistance with the GWU-QCD code, and Jacob Finkenrath and Michael Engelhardt for
helpful discussions. We are grateful to GWU IMPACT (Institute for Massively Parallel Applications and Computing Technologies) for CPU time and to the Columbian College
for logistical support of our GPU cluster. This work was supported in part by DoE grant DE-FG02-95ER40907 and NSF CAREER grant PHY-1151648.

\newpage \samepage

\bibliographystyle{JHEP}
\bibliography{my-references}

\providecommand{\href}[2]{#2}\begingroup\raggedright\begin{thebibliography}{10}

\bibitem{Schmiedmayer:1991zz}
J.~Schmiedmayer, P.~Riehs, J.~A. Harvey, and N.~W. Hill, {\it {Measurement of
  the electric polarizability of the neutron}},  {\em Phys.Rev.Lett.} {\bf 66}
  (1991) 1015--1018.

\bibitem{Kossert:2002ws}
K.~Kossert, M.~Camen, F.~Wissmann, J.~Ahrens, J.~Annand, {\em et.~al.}, {\it
  {Quasifree Compton scattering and the polarizabilities of the neutron}},
  {\em Eur.Phys.J.} {\bf A16} (2003) 259--273,
  [\href{http://xxx.lanl.gov/abs/nucl-ex/0210020}{{\tt nucl-ex/0210020}}].

\bibitem{Alexandru:2009id}
A.~Alexandru and F.~X. Lee, {\it {Neutron electric polarizability}},  {\em PoS}
  {\bf LAT2009} (2009) 144, [\href{http://xxx.lanl.gov/abs/0911.2520}{{\tt
  arXiv:0911.2520}}].

\bibitem{Lujan:2011ue}
M.~Lujan, A.~Alexandru, and F.~Lee, {\it {Electric polarizability of hadrons
  with overlap fermions on multi-GPUs}},  {\em PoS} {\bf LATTICE2011} (2011)
  165, [\href{http://xxx.lanl.gov/abs/1111.6288}{{\tt arXiv:1111.6288}}].

\bibitem{Alexandru:2010dx}
A.~Alexandru and F.~Lee, {\it {Hadron electric polarizability -- finite volume
  corrections}},  {\em PoS} {\bf LATTICE2010} (2010) 131,
  [\href{http://xxx.lanl.gov/abs/1011.6309}{{\tt arXiv:1011.6309}}].

\bibitem{Engelhardt:2010tm}
M.~Engelhardt, {\it {Progress toward the chiral regime in lattice QCD
  calculations of the neutron electric polarizability}},  {\em PoS} {\bf
  LAT2009} (2009) 128, [\href{http://xxx.lanl.gov/abs/1001.5044}{{\tt
  arXiv:1001.5044}}].

\bibitem{Bazavov:2009bb}
A.~Bazavov, D.~Toussaint, C.~Bernard, J.~Laiho, C.~DeTar, {\em et.~al.}, {\it
  {Nonperturbative QCD simulations with 2+1 flavors of improved staggered
  quarks}},  {\em Rev.Mod.Phys.} {\bf 82} (2010) 1349--1417,
  [\href{http://xxx.lanl.gov/abs/0903.3598}{{\tt arXiv:0903.3598}}].

\bibitem{Hasenfratz:2007rf}
A.~Hasenfratz, R.~Hoffmann, and S.~Schaefer, {\it {Hypercubic smeared links for
  dynamical fermions}},  {\em JHEP} {\bf 0705} (2007) 029,
  [\href{http://xxx.lanl.gov/abs/hep-lat/0702028}{{\tt hep-lat/0702028}}].

\bibitem{Sommer:1993ce}
R.~Sommer, {\it {A New way to set the energy scale in lattice gauge theories
  and its applications to the static force and alpha-s in SU(2) Yang-Mills
  theory}},  {\em Nucl.Phys.} {\bf B411} (1994) 839--854,
  [\href{http://xxx.lanl.gov/abs/hep-lat/9310022}{{\tt hep-lat/9310022}}].

\bibitem{Pelissier:2012pi}
C.~Pelissier and A.~Alexandru, {\it {Resonance parameters of the rho-meson from
  asymmetrical lattices}},  \href{http://xxx.lanl.gov/abs/1211.0092}{{\tt
  arXiv:1211.0092}}.

\bibitem{Detmold:2009dx}
W.~Detmold, B.~C. Tiburzi, and A.~Walker-Loud, {\it {Extracting Electric
  Polarizabilities from Lattice QCD}},  {\em Phys.Rev.} {\bf D79} (2009)
  094505, [\href{http://xxx.lanl.gov/abs/0904.1586}{{\tt arXiv:0904.1586}}].

\bibitem{Alexandru:2008sj}
A.~Alexandru and F.~X. Lee, {\it {The Background field method on the lattice}},
   {\em PoS} {\bf LATTICE2008} (2008) 145,
  [\href{http://xxx.lanl.gov/abs/0810.2833}{{\tt arXiv:0810.2833}}].

\bibitem{Rubinstein:1995hc}
H.~Rubinstein, S.~Solomon, and T.~Wittlich, {\it {Dependence of lattice hadron
  masses on external magnetic fields}},  {\em Nucl.Phys.} {\bf B457} (1995)
  577--593, [\href{http://xxx.lanl.gov/abs/hep-lat/9501001}{{\tt
  hep-lat/9501001}}].

\bibitem{Liu:2012gm}
Q.~Liu, N.~H. Christ, and C.~Jung, {\it {Light Quark Mass Reweighting}},
  \href{http://xxx.lanl.gov/abs/1206.0080}{{\tt arXiv:1206.0080}}.

\bibitem{Hasenfratz:2008fg}
A.~Hasenfratz, R.~Hoffmann, and S.~Schaefer, {\it {Reweighting towards the
  chiral limit}},  {\em Phys.Rev.} {\bf D78} (2008) 014515,
  [\href{http://xxx.lanl.gov/abs/0805.2369}{{\tt arXiv:0805.2369}}].

\bibitem{Ishikawa:2012ix}
T.~Ishikawa, T.~Blum, M.~Hayakawa, T.~Izubuchi, C.~Jung, {\em et.~al.}, {\it
  {Full QED+QCD low-energy constants through reweighting}},  {\em
  Phys.Rev.Lett.} {\bf 109} (2012) 072002,
  [\href{http://xxx.lanl.gov/abs/1202.6018}{{\tt arXiv:1202.6018}}].

\bibitem{Ohki:2009mt}
H.~Ohki, S.~Aoki, H.~Fukaya, S.~Hashimoto, T.~Kaneko, {\em et.~al.}, {\it
  {Nucleon sigma term and strange quark content in 2+1-flavor QCD with
  dynamical overlap fermions}},  {\em PoS} {\bf LAT2009} (2009) 124,
  [\href{http://xxx.lanl.gov/abs/0910.3271}{{\tt arXiv:0910.3271}}].

\bibitem{Finkenrath:2012cz}
J.~Finkenrath, F.~Knechtli, and B.~Leder, {\it {Application of Domain
  Decomposition to the Evaluation of Fermion Determinant Ratios}},
  \href{http://xxx.lanl.gov/abs/1211.1214}{{\tt arXiv:1211.1214}}.

\bibitem{Alexandru:2010yb}
A.~Alexandru and U.~Wenger, {\it {QCD at non-zero density and canonical
  partition functions with Wilson fermions}},  {\em Phys.Rev.} {\bf D83} (2011)
  034502, [\href{http://xxx.lanl.gov/abs/1009.2197}{{\tt arXiv:1009.2197}}].

\bibitem{Luscher:2008tw}
M.~Luscher and F.~Palombi, {\it {Fluctuations and reweighting of the quark
  determinant on large lattices}},  {\em PoS} {\bf LATTICE2008} (2008) 049,
  [\href{http://xxx.lanl.gov/abs/0810.0946}{{\tt arXiv:0810.0946}}].

\bibitem{Alexandru:2002jr}
A.~Alexandru and A.~Hasenfratz, {\it {Partial global stochastic metropolis
  update for dynamical smeared link fermions}},  {\em Phys.Rev.} {\bf D66}
  (2002) 094502, [\href{http://xxx.lanl.gov/abs/hep-lat/0207014}{{\tt
  hep-lat/0207014}}].

\bibitem{Thron:1997iy}
C.~Thron, S.~Dong, K.~Liu, and H.~Ying, {\it {Pade - Z(2) estimator of
  determinants}},  {\em Phys.Rev.} {\bf D57} (1998) 1642--1653,
  [\href{http://xxx.lanl.gov/abs/hep-lat/9707001}{{\tt hep-lat/9707001}}].

\end{thebibliography}\endgroup

\end{document}